\documentclass[journal]{IEEEtran}
\usepackage{multirow}
\usepackage{subfigure}
\usepackage{graphicx}
\usepackage{amsmath}
\usepackage{amsfonts}
\usepackage{algorithm}
\usepackage{algorithmic}
\usepackage{makecell}
\usepackage{booktabs}
\usepackage{autobreak}
\setcellgapes{3pt}

\setcellgapes{3pt}

\usepackage[justification=centering]{caption}
\usepackage{color}

\usepackage{cite}
\usepackage{verbatim}

\ifCLASSINFOpdf
\else
\fi
\begin{document}
%
\title{Attention-based UAV Trajectory Optimization for Wireless Power Transfer-assisted IoT Systems}
\author{Li Dong, Feibo Jiang, \textit{Senior Member, IEEE}, Yubo Peng.

\thanks{This work was supported in part by the National Natural Science Foundation of China under Grant 41904127, Grant 41604117 and Grant 62132004, in part by the Hunan Provincial Natural Science Foundation of China under Grant 2024JJ5270, in part by the Open Project of Xiangjiang Laboratory under Grant 22XJ03011, Grant XJ2023001 and Grant XJ2022001, in part by the Scientific Research Fund of Hunan Provincial Education Department under Grant 22B0663, in part	by the Changsha Natural Science Foundation under Grant kq2402098, and Grant kq2402162, and in part by Qiyuan Lab Innovation Fund under Grant 2022-JCJQ-LA-001-088. (Corresponding author: Feibo Jiang)}
\thanks{Li Dong is with the School of Computer Science, Hunan University of Technology and Business, Changsha 410205, China, and also with the Xiangjiang Laboratory, Changsha 410205, China (e-mail: Dlj2017@hunnu.edu.cn).}
\thanks{Feibo Jiang is with the Hunan Provincial Key Laboratory of Intelligent
	Computing and Language Information Processing, Hunan Normal
	University, Changsha 410081, China (e-mail: jiangfb@hunnu.edu.cn).}
\thanks{Yubo Peng is with the School of Intelligent Software and Engineering,
Nanjing University, Suzhou 215163, China (e-mail: pengyubo@
	hunnu.edu.cn).}


}
\markboth{Submitted for Review}%
{Shell \MakeLowercase{\textit{et al.}}: Bare Demo of IEEEtran.cls for IEEE Journals}
%



\maketitle
\begin{abstract}
\textcolor{black}{Unmanned Aerial Vehicles (UAVs) in Wireless Power Transfer (WPT)-assisted Internet of Things (IoT) systems face the following challenges: limited resources and suboptimal trajectory planning. Reinforcement learning-based trajectory planning schemes face issues of low search efficiency and learning instability when optimizing large-scale systems. To address these issues, we present an Attention-based UAV Trajectory Optimization (AUTO) framework based on the graph transformer, which consists of an Attention Trajectory Optimization Model (ATOM) and a Trajectory lEarNing Method based on Actor-critic (TENMA). In ATOM, a graph encoder is used to calculate the self-attention characteristics of all IoTDs, and a trajectory decoder is developed to optimize the number and trajectories of UAVs. TENMA then trains the ATOM using an improved Actor-Critic method, in which the real reward of the system is applied as the baseline to reduce variances in the critic network. This method is suitable for high-quality and large-scale multi-UAV trajectory planning. Finally, we develop numerous experiments, including a hardware experiment in the field case, to verify the feasibility and efficiency of the AUTO framework.}

\end{abstract}


%
\IEEEpeerreviewmaketitle

\section{Introduction}
\label{sec:introduction}

%
%
%
%

With the advancement of 5G, the Internet of Things (IoT) has become widely used in a variety of fields, including environmental monitoring, healthcare, and industry 4.0, among others. However, due to limited transmitting power and battery capacity, Internet of Things Devices (IoTDs) perform poorly in long-distance communication. Furthermore, when IoTDs are positioned in remote places with limited wireless coverage and battery power, charging IoTDs and transferring sensory data from the IoTDs to the remote data center are challenging tasks.

\textcolor{black}{Fortunately, Unmanned Aerial Vehicles (UAVs) and Wireless Power Transfer (WPT) can be integrated into these IoT systems, enabling UAVs to wirelessly transfer power to IoTDs \cite{messaoudi2023survey}. Before collecting the sensory data of an IoTD, the UAV needs to charge the IoTD so that the IoTD has enough energy to transfer its sensory data to the UAV. 
Thus, UAVs are deployed in the IoT system as mobile chargers to transfer power to IoTDs and as the data collectors for sensory data collection. 
Compared to terrestrial data collection systems, UAVs can more efficiently cover large areas and fly close to IoTDs for data collection and energy transfer. This reduces data transmission latency, alleviates communication burdens, and enhances the efficiency of WPT
\cite{bouzid20235g}.}
Hence, mobility management plays a key role for UAVs in WPT-assisted IoT systems, and the shoddy design of UAV trajectory will lead to not only the waste of energy but also service delays. Efficient mobility management is demanded in trajectory design, especially considering a swarm of UAVs \cite{dong2024deep,10726905,jiang2024large}.

There are many previous trajectory design algorithms have been proposed. 
\textcolor{black}{Messaoudi et al. \cite{messaoudi2024ugv} proposed a collaborative system based on UAVs and Unmanned Ground Vehicles (UGVs) for the data collection of IoT devices. The trajectory control of UGVs and UAVs was optimized using a multi-agent Reinforcement Learning (RL) approach. 
Lu et al. \cite{lu2021covertness} introduced a WPT system utilizing UAVs, which first wirelessly charge energy-constrained IoT devices and then enable these devices to opportunistically send collected data to the UAV. 
Oubbati et al. \cite{oubbati2022synchronizing} proposed a multi-agent deep RL method called TEAM to optimize UAV trajectories and resource allocation while minimizing UAV energy consumption. Zhu et al. \cite{zhu2022uav} introduced a machine learning algorithm based on Transformer and Weighted A*, named TWA, to address UAV trajectory optimization in UAV-aided IoT networks. 
Zhu et al. \cite{10224843} proposed an Attention-Reinforced Learning Scheme for optimizing the trajectory of UAVs in large-scale and low-power data collection tasks.}

\begin{table}[]
	\caption{Comparison with previous works}
	\centering
	\setlength{\tabcolsep}{5mm}
	\renewcommand\arraystretch{1.25}
	\begin{tabular}{|c|c|c|c|c|}
	\hline
Work & UAV & WPT & Attention & RL \\
	\hline
\cite{messaoudi2024ugv} & $\checkmark$ & $\checkmark$ &  & $\checkmark$ \\
	\hline
	\cite{lu2021covertness} & $\checkmark$ & $\checkmark$ &   &   \\
	\hline
	\cite{oubbati2022synchronizing} & $\checkmark$ & $\checkmark$ &   & $\checkmark$ \\
	\hline
	\cite{zhu2022uav} & $\checkmark$ &   & $\checkmark$ &   \\
	\hline
	\cite{10224843} & $\checkmark$ &   & $\checkmark$ & $\checkmark$ \\
	\hline
	AUTO & $\checkmark$ & $\checkmark$ & $\checkmark$ & $\checkmark$ \\
	\hline
\end{tabular}
	\label{tab:relatedwork}
\end{table}

Although high mobility is an impressive feature of UAVs, the battery and data storage capacity of UAVs always limits the data collection tasks for all IoTDs in the system. Therefore, joint resource and mobility management is essential to UAV swarms for the WPT-assisted IoT system \cite{jiang2020ai,jiang2021distributed,10638533}. 
In this paper, we propose a novel Attention-based UAV Trajectory Optimization (AUTO) framework, in which an Attention Trajectory Optimization Model (ATOM) is used to optimize the number and trajectories of UAVs, and a Trajectory lEarNing Method based on Actor-critic (TENMA) is applied to train the ATOM model. 

\textcolor{black}{In Table \ref{tab:relatedwork}, we compare our AUTO framework with existing works, focusing on UAV, WPT, attention mechanisms, and RL. It is evident that most of the listed works consider the system from only two or three perspectives. However, the aforementioned studies do not specifically investigate the potential of integrating the attention and RL in UAV and WPT-assisted IoT systems.
Hence, unlike existing works, the contributions of the study are summarized as follows: }
\begin{enumerate}

	\item 
\textcolor{black}{	\emph{High-precision graph encoding:} In ATOM, the IoT system is mathematically expressed as a graph structure and a graph encoder is presented to extract the self-attention features of all IoTDs precisely. The novel learnable graph embedding layer and graph pooling layer are introduced to the graph encoder for enhancing self-attention features and guiding the trajectory planning.}

	\item
\textcolor{black}{	\emph{High-quality trajectory decoding:} The trajectory decoder followed by the graph encoder in ATOM is utilized to generate trajectories of UAVs. We decode the self-attention features of each IoTD and the whole graph by the proposed alignment vector and context vector, and the quantity and trajectories of UAVs are optimized according to self-attention features and remaining battery and storage capacity.}

	\item 
\textcolor{black}{	 \emph{Efficient and stable Actor-Critic learning:} The ATOM model is trained by TENMA. Specifically, a critic network is applied to evaluate the generated trajectories, and the ATOM model is introduced as the actor network to produce the trajectories. Moreover, the real reward of the system is used as the baseline to reduce the variance of the critic network and enhance the stabilization and generalization of TENMA.}

\end{enumerate}

This paper is organized as follows: 
Section \ref{sec:system_model} describes the system model and problem formulation of the WPT-assisted IoT system. 
Section \ref{sec:UFO} describes the principle of the AUTO framework in details. 
Section \ref{sec:experiments} illustrates the experiment results. Finally, Section \ref{sec:conclusion} concludes the whole paper.

\section{System Model and Problem Formulation }
\label{sec:system_model}

\begin{figure}[htbp]
\centering
\includegraphics[width=9cm]{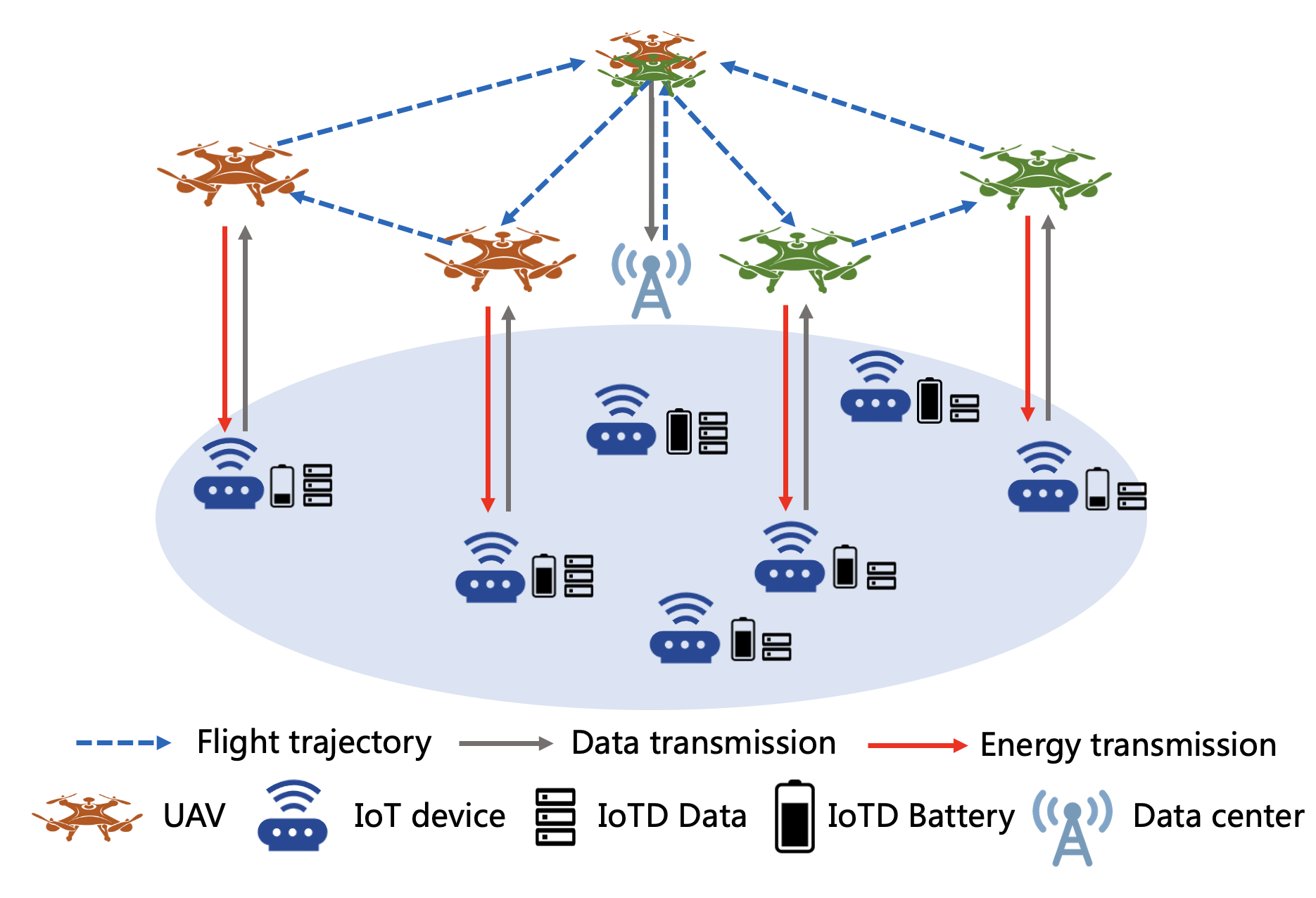}
\caption{The WPT-assisted IoT system.}
\label{fig:scenario}
\end{figure}

As illustrated in Fig. \ref{fig:scenario}, the WPT-assisted IoT system consists of $N$ IoTDs, a data center, and $m$ UAVs with half-duplex access points, which can transmit power to the IoTDs and collect data from IoTDs by Time Division Duplexing (TDD) mode.
The $N$ IoTDs are denoted as a set of $\mathcal{N}=\left\{ 1,2,...,N \right\}$. We assumed that the position ($x_i, y_i$) of the $i$-th IoTD is fixed and known, and the $i$-th IoTD has $D_{i}$ data to be collected. 
The set of $m$ UAVs is denoted as $\mathcal{M}=\left\{ 1,2,...,m \right\}$, and each UAV has limited data storage capacity $C_{\max}$ and energy capacity $E_{\max}$. The flight height of the UAV is set to $H^F$. Each UAV can only collect data from one IoTD at one time, so the association $a_{ij}$ at the $t$-th time step can be expressed by:
\begin{equation}
\label{eq:c_uav_collect}
a_{ij}[t]=\left\{0,1\right\}, \forall i\in\mathcal{N}, \forall j\in\mathcal{M}
\end{equation}
where $a_{ij}[t]=1$ means the $j$-th UAV is collecting the data from the $i$-th IoTD at the $t$-th time step, and $a_{ij}[t]=0$ otherwise.

\subsection{Trajectory Model}

In the proposed system, each UAV flies straightly from
one IoTD to another. The $j$-th UAV takes off from the
data center at a fixed location $r_j[0]=(0,0,H)$, and flies to the IoTD one by one, and hovers over each IoTD to collect data. The $j$-th
UAV completes the data collection task according to a predetermined flight trajectory and returns to the same data center after one flying cycle. Hence, we have
\begin{equation}
\label{eq:c_return}
r_j[s_j]=r_j[0],\forall j \in\mathcal{M}
\end{equation}
where $r_j[t]$ denotes the $t$-th hover point on the flight trajectory of the $j$-th UAV, $t \in\mathcal{T}_j=\left\{ 1,2,...,s_j \right\}$, and the $j$-th UAV serves total $s_j$ IoTDs. Hence, the $j$-th UAV has $s_j$ hover points. 
Assume that the UAV serves the $i$-th IoTD only once, then one has 
\begin{equation}
\label{eq:c_iot_transfer}
\sum_{j=1}^{m}\sum_{t=1}^{s_j} a_{ij}[t]=1, \forall i \in\mathcal{N}.
\end{equation}

Since each UAV can fly from one hover point to another point in a straight line, then the flight time of the $j$-th UAV can be expressed by:
\begin{equation}
\label{eq:uav_flight_time}
T_{j}^{F}=\sum_{t=1}^{s_j}\frac{\left\|r_j[t]-r_j[t-1]\right\|_2}{v},  \forall j \in \mathcal{M}
\end{equation}
where $\left\|r_j[t+1]-r_j[t]\right\|_2$ is the Euclidean distance between hover points $r_j[t+1]$ and $r_j[t]$, and $v$ is the flight velocity of the UAV, which is a constant value.

\subsection{Data Collection Model}

We assume that IoTDs can be wirelessly charged by the UAV before transmitting the data to the UAV. The whole process can be divided into the WPT stage and data transmission stage.

In the WPT stage, the UAV can transmit the energy wirelessly via Radio Frequency (RF) technologies with a fixed transmit power $P^{T}$. The power received at the $i$-th IoTD from the $j$-th UAV is denoted as $P^{R}_{ij}$, which can be calculated by
\begin{equation}
\label{eq:eh_receive}
P^{R}_{ij}=|g^{D}_{ij}|^2 P^{T}  
\end{equation}
where $|g^{D}_{ij}|^2$ denotes the downlink power gain from the $j$-th UAV to the $i$-th IoTD.

Assume that $\eta^{L}$ is the constant attenuation parameter in the linear energy harvesting model. 
Then, the received energy of the $i$-th IoTD from the $j$-th UAV can be given by
\begin{equation}
\label{eq:linear_eh_model}
E^{R}_{ij}=\eta^{L}P^{R}_{ij} T_{ij}^{E}
\end{equation}
where $T_{ij}^{E}$ is the energy harvesting time of the $i$-th IoTD from the $j$-th UAV.

In the data transmission stage, the uploading data rate of the  $i$-th IoTD to the $j$-th UAV can be given by
\begin{equation}
\label{eq:jiang3}
R_{ij}=B\log_2(1+\frac{|g^{U}_{ij}|^2 E^{R}_{ij}}{\sigma^2 T^{C}_{ij}})
\end{equation}
where $|g^{U}_{ij}|^2=|g^{D}_{ij}|^2$ denotes the uplink power gain, $B$ is the bandwidth, $\sigma^2$ is Gaussian white noise power, and $T^{C}_{ij}$ is the data collection time from $i$-th IoTD to the $j$-th UAV. 

To ensure that IoTDs can successfully upload their data $D_{i}$ to the UAVs, one has
\begin{equation}
\label{eq:c_eh}
T^{C}_{ij}B\log_2(1+\frac{|g^{U}_{ij}|^2 E^{R}_{ij}}{\sigma^2 T^{C}_{ij}}) \geq D_{i}, \forall j\in \mathcal{M}.
\end{equation}

%

\subsection{Energy Consumption Model}

Assuming that the flight energy consumption of the $j$-th UAV is given as
\begin{equation}
\label{eq:uav_flight_energy}
E_j^F=P^FT_j^F
\end{equation}
where $P^F$ is the flight power of the UAV. 
We also assume that the power consumption is $P^{H}$ when the $j$-th UAV hovers above the IoTD, then one has
\begin{equation}
 \label{eq:jiang4}
E_j^T=({P^{T}}+P^{H}) T^{T}_j
\end{equation}
where  $T^{T}_j$ is the power transfer time of the $j$-th UAV. 
Next, when the $j$-th UAV hovers, the energy consumption of data collection and UAV hovering is calculated as
\begin{equation}
\label{eq:uav_hover_energy}
E_{j}^{C}=(P^{C}+P^{H}) T^{C}_j
\end{equation}
where $T_j^{C}$ is the data collection time of the $j$-th UAV. $P^{C}$ is the data collection power of the UAV.
Therefore, the total energy consumption of the $j$-th UAV can be given as
\begin{equation}
\label{eq:total_energy}
E_j=E_j^F+E_{j}^{C}+E_j^T, \forall j\in\mathcal{M}.
\end{equation}

Due to the limited data storage capacity and energy capacity of UAVs, it is required that the total energy consumption of the UAV does not exceed its energy capacity $E_{\max}$ and all collected data does not exceed the storage capacity $C_{\max}$. Therefore, these inequalities need to be satisfied
\begin{equation}
\label{eq:c_battery}
E_j\leq E_{\max}, \forall j\in\mathcal{M},
\end{equation}
\begin{equation}
\label{eq:c_capacity}
\sum_{i=1}^N \sum_{t=1}^{s_j}a_{ij}[t]D_{i}\leq C_{\max}, \forall j\in\mathcal{M}.
\end{equation}

\subsection{Problem Formulation}
We aim to minimize the energy consumption of all UAVs, by jointly optimizing the number and trajectories of UAVs, the user association, the energy harvesting time, and the data collection time.
The optimization problem can be mathematically formulated by
\begin{equation}
	\label{eq:problem}
	\begin{aligned}
		P0:\min_{\mathcal{A},m,\mathcal{R}, \mathcal{T}^{E}, \mathcal{T}^{C}}\sum_{j=1}^m E_j \\
		\text{s.t.}~  \eqref{eq:c_uav_collect}, \eqref{eq:c_return},\eqref{eq:c_iot_transfer}, \eqref{eq:c_eh}, \eqref{eq:c_battery},\eqref{eq:c_capacity}
	\end{aligned}
\end{equation}
where $\mathcal{A}=\left\{a_{ij}[t], \forall i \in \mathcal{N}, j \in \mathcal{M},  t \in \mathcal{T}_j \right\}$ represents the association between UAVs and IoTDs. $m$ is the optimal number of UAVs for data collection. $\mathcal{R}=\left\{r_j[t], \forall j \in \mathcal{M},  t \in \mathcal{T}_j \right\}$ represents the visiting order of hover points for UAVs. $\mathcal{T}^{E}=\left\{T_{ij}^{E}, \forall i \in \mathcal{N}, j \in \mathcal{M}, t \in \mathcal{T}_j \right\}$ represents the set of energy harvesting time and $\mathcal{T}^{C}=\left\{T_{ij}^{C}, \forall i \in \mathcal{N}, j \in \mathcal{M}, t \in \mathcal{T}_j \right\}$ represents the set of collecting data time. 

\subsection{Problem Decomposition}
Since the energy harvesting time and the data collection time of each IoTD are independent of trajectories of UAVs, the original problem $P0$ can be decomposed into two sub-problems: time allocation problem $P1$ and trajectory optimization problem $P2$. Problem $P1$ can be described as follows:
%
%
%
\begin{equation}
\label{eq:jiang5}
\begin{aligned}
P1:\min_{\mathcal{T}^{C}, \mathcal{T}^{E}}\sum_{j=1}^{m} (E^H_j+E^T_j) \\
\text{s.t.}~ \eqref{eq:c_eh}.\ \ \ \ \ \ \ \ \ \ \ \ 
\end{aligned}
\end{equation}
%
%
%
%
%
%
%

One can see that Problem $P1$ is a convex optimization problem, 
 which can be solved by applying Karush-Kuhn-Tucker (KKT) conditions \cite{2012-kkt}. The optimal solutions $\left\{ {T^{C}_{ij}}^{*}, {T^{E}_{ij}}^{*} \right\}$ can be calculated as follows\cite{2020-aoi-eh}:
\begin{equation}
\label{eq:t_cd_solution}
{T^{C}_{ij}}^{*}=\frac{\ln(2\cdot D_{i})}{\mathcal{W}(\frac{{|g^{D}_{ij}|^2 \phi(P^{R}_{ij})-{\sigma^2}}}{e \cdot {\sigma^2}})+1}, \forall i \in \mathcal{N}, \forall j\in \mathcal{M}
\end{equation}
where $\mathcal{W}(\cdot)$ is the Lambert W function\cite{2020-aoi-eh} and
\begin{equation}
\label{eq:t_eh_solution}
{T^{E}_{ij}}^{*}=\frac{{T^{C}_{ij}}^{*}(2^{\frac{D_{i}}{{T^{C}_{ij}}^{*}}}-1){\sigma^2}}{{{|g^{D}_{ij}|^2 \phi(P^{R}_{ij})}}}, \forall i \in \mathcal{N},  \forall j\in \mathcal{M}.
\end{equation}

After obtaining the optimal $\left\{{{T^{C}_{ij}}^{*}, {T^{E}_{ij}}^{*}}\right\}$, Problem $P2$ can be described as follows:
\begin{equation}
\label{eq:jiang6}
\begin{aligned}
P2:\min_{\mathcal{A},m,\mathcal{R}}\sum_{j=1}^m E^F_j \\
\text{s.t.}~      
\eqref{eq:c_uav_collect}, \eqref{eq:c_return},\eqref{eq:c_iot_transfer}, 
\eqref{eq:c_battery},\eqref{eq:c_capacity}.
\end{aligned}
\end{equation}

\section{Attention-based UAV Trajectory Optimization Framework}
\label{sec:UFO}

Problem $P2$ can be formulated as a well-known combinatorial optimization problem called Capacitated Vehicle Routing Problem (CVRP), which is NP-hard and difficult to be solved. 
Self-attention can encode the whole system information (e.g., the quantity, location and data size of IoTDs) to a self-attention feature matrix, and make a global decision based on it. Hence, we propose the AUTO framework based on the graph transformer to solve Problem $P2$. 


\subsection{AUTO Framework Overview}
In the AUTO framework, we propose a novel ATOM model to optimize the quantity, trajectory, and association by the customized graph self-attention model. Moreover, we propose a new TENMA method to train the ATOM model. 

The workflow of the AUTO framework is presented in \textbf{Algorithm \ref{alg:overview}}. 
We initialize the parameter of ATOM model $\theta_\pi$ randomly at the beginning. 
Then, in the training stage, we utilize TENMA method to train the ATOM model. 
In the inference stage, 
the self-attention features of all IoTDs $\left\{\textbf{h}_1^L,...,\textbf{h}_N^L\right\}$ and the graph feature $\textbf{h}_{sa}$ are obtained by the graph encoder. Next, a solution that contains the total trajectory $\boldsymbol{\pi}$ and quantity $m$ of UAVs are obtained by trajectory decoder. The solution also meets the constraints of storage and battery capacity. Finally, we split the total trajectory $\boldsymbol{\pi}$ into trajectories $\left\{\boldsymbol{\pi}_{(1)},...,\boldsymbol{\pi}_{(m)} \right\}$ of all UAVs. In the $j$-th UAV trajectory, the visiting order is denoted as $\boldsymbol{\pi}_{(j)}=[\pi_{j,1},...,\pi_{j,s_j}], $ and $\pi_{j,t}$ is the index of the IoTD whose data is collected by the $j$-th UAV in the $t$-th time step. 
Therefore, all the association \textcolor[rgb]{0,0,0}{$a_{ij}[t]$} and visiting order \textcolor[rgb]{0,0,0}{$r_{j}[t]$}  are solved in the inference stage of \textbf{Algorithm \ref{alg:overview}}.

\begin{algorithm}
\caption{AUTO framework}
\textbf{Input:} $C_{\max}$, $E_{\max}$.

\textbf{Output:} \textcolor[rgb]{0,0,0}{$a_{ij}[t]$}, $m$, \textcolor[rgb]{0,0,0}{$r_{j}[t]$}. 

\begin{algorithmic}[1]

\STATE Initialize parameters $\theta_\pi$ of the ATOM model.
\STATE Initialize $a_{ij}[t]=0, r_{j}[t]=0$.

\textbf{Training stage}

\STATE Train ATOM and update $\theta_\pi$ by TENMA method.

\textbf{Inference stage}

\STATE Calculate the $\textbf{h}_{sa}$ and $\left\{\textbf{h}_1^L,...,\textbf{h}_N^L\right\}$ by Graph Encoder. 
\STATE Calculate $\boldsymbol{\pi}$ and $m$ by Trajectory Decoder. 
\STATE Split $\boldsymbol{\pi}$ into $\left\{\boldsymbol{\pi}_{(1)},...,\boldsymbol{\pi}_{(m)} \right\}$. 
\FOR {$j=1,...,m$}
\FOR {$t=1,...,s_j$}
\STATE $r_j[t]=\pi_{j,t}$. 
\STATE $a_{\pi_{j,t}, j}[t]=1$. 
\ENDFOR
\ENDFOR

\end{algorithmic}
\label{alg:overview}
\end{algorithm}


\subsection{Graph Encoder}
\label{sec:encoder}

The proposed IoT system can be mathematically expressed as a graph structure pair: $\mathcal{G}=(\mathcal{X},\mathcal{E})$, where $\mathcal{X}$ is the set of IoTD node, and $\mathcal{E}$ is the set of edges. For one IoTD, the neighborhoods in the communication range all have edges connected to it. 
We propose a novel graph encoder to extract different self-attention features from IoTDs and the whole graph, which can help decoder to generate trajectories of UAVs with minimum energy consumption. The graph encoder contains a graph embedding layer, $L$ self-attention layers and a graph pooling layer. Each self-attention layer contains a multi-head self-attention (MSA) sublayer, and a fully connected (FC) sublayer. Moreover, residual connections are introduced after every MSA sublayer and fully connected sublayer, and normalization operators are introduced before every MSA sublayer and fully connected sublayer \cite{2017transformer}. 
\textcolor{black}{The proposed graph encoder is illustrated in Fig. \ref{fig:encoder}}. 
The detailed procedure of the graph encoder can be described as follows:
\begin{figure}[htbp]
\centering
\includegraphics[width=9cm]{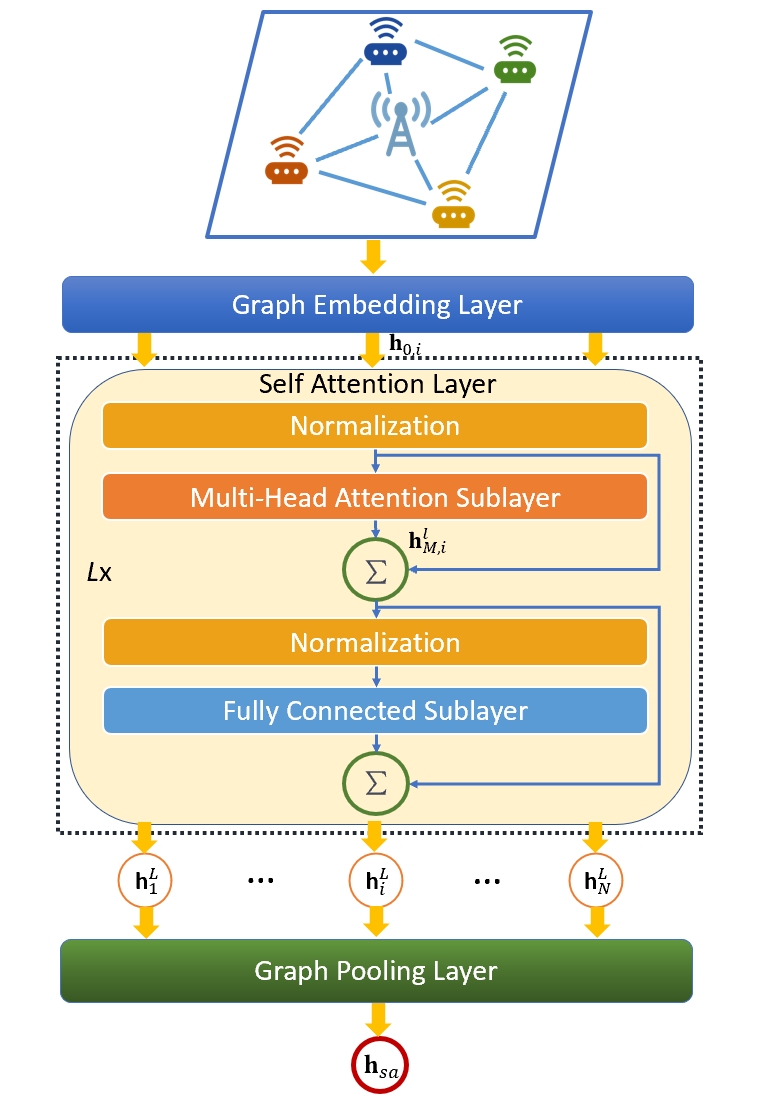}
\caption{Structure of the graph encoder.}
\label{fig:encoder}
\end{figure}
\subsubsection{Graph Embedding Calculation}

\textcolor{black}{We define the IoTD information contains the coordinate $\{x^c_i, y^c_i\}$ of the $i$-th IoTD and sensory data $D_{i}$ generated from the $i$-th IoTD, which is denoted as $\textbf{X}_i=[x^c_i, y^c_i, D_{i}]$ and the information of all IoTDs is denoted as $\mathcal{X}=\left\{\textbf{X}_i, \forall i \in \mathcal{N}\right\}$. The graph embedding layer is applied to preprocess the graph information $\mathbf{h}_{0,i}$ of the $i$-th IoTD.}

\subsubsection{Self Attention Calculation}

\textcolor{black}{We introduce self attention layer to calculate self-attention features of all IoTDs.}

\textcolor{black}{First, the graph encoder maps the embedding information $\textbf{h}_{0,i}$ of the $i$-th IoTD  to query $\textbf{Q}_i$, key $\textbf{K}_i$ and value $\textbf{V}_i$ with learnable matrices, respectively. }

\textcolor{black}{The attention scores are computed as the dot product of the query $\textbf{Q}_i$ and key $\textbf{K}_j$, normalized by the square root of the dimension of the key vectors, followed by a softmax function to obtain the attention weights. These weights are then used to compute a weighted sum of the value $\textbf{V}_j$, producing the self-attention features $\textbf{Z}_i$ of the $i$-th IoTD for each head.}

\textcolor{black}{The outputs of all attention heads are concatenated and linearly transformed to form the MSA feature $\textbf{h}^l_{M,i}$ of the $i$-th IoTD at the $l$-th self-attention layer. The multi-head mechanism allows the self-attention layers to jointly attend to extracted information from different representation subspaces at different IoTDs \cite{2017transformer}.} The output of the $i$-th IoTD in the $l$-th self attention layer is
\begin{equation}
	\label{eq:a3}
	\textbf{h}^l_i=\operatorname{FC}(\textbf{h}^l_{M,i})+\textbf{h}^l_{M,i}
\end{equation}
where $\operatorname{FC}(\cdot)$ means the feed-forward operator of the fully connected layer.

\subsubsection{Graph Pooling Calculation}
\textcolor[rgb]{0,0,0}{After calculating the self-attention features of all IoTDs, we need to merge all attention features to a global attention feature. However, simply averaging the features may produce an unrecoverable loss of IoTDs information\cite{2014learnable-pool}.} 
\textcolor[rgb]{0,0,0}{
Therefore, we propose a graph pooling layer to merge attention features of all IoTDs $\left\{\textbf{h}_1^L,...,\textbf{h}_N^L\right\}$ to a graph feature $\textbf{h}_{sa}$, which can be given as}
\begin{align}
	\mathbf{h}_{sa} &= \operatorname{Mean}\left(\left\{\mathbf{W}_{G,i} \cdot \operatorname{Concat}\left(\mathbf{h}^L_{i}, \mathbf{h}^L_{j}\right), \right.\right. \notag \\
	& \qquad \left.\left. \forall\ i\in\mathcal{N};\ j\in\mathcal{N}(i) \right\}\right)
	\label{eq:a2}
\end{align}
where $\mathbf{h}_{L,i}$ and $\mathbf{h}_{L,j}$ are the outputs of the $L$-th self attention layer, $\mathbf{W}_{G,i}$ is the learnable matrix. $\operatorname{Concat}(\cdot)$ means the concatenation of two vectors.

\subsection{Trajectory Decoder}
\label{sec:decoder}

In Problem $P2$, the aim is to minimize the system energy consumption by jointly optimizing the number and the trajectories of UAVs. 
The UAV takes off from the data center, collects data from IoTDs one by one, and returns to the data center. Then, the next UAV repeats the steps until all trajectories are generated. The procedure of the trajectory planning in trajectory decoder is illustrated in Fig. \ref{fig:decoder}. 
 The detailed process of the trajectory decoder is explained as follows: 

\begin{figure*}[htbp]
\centering
\includegraphics[width=14cm]{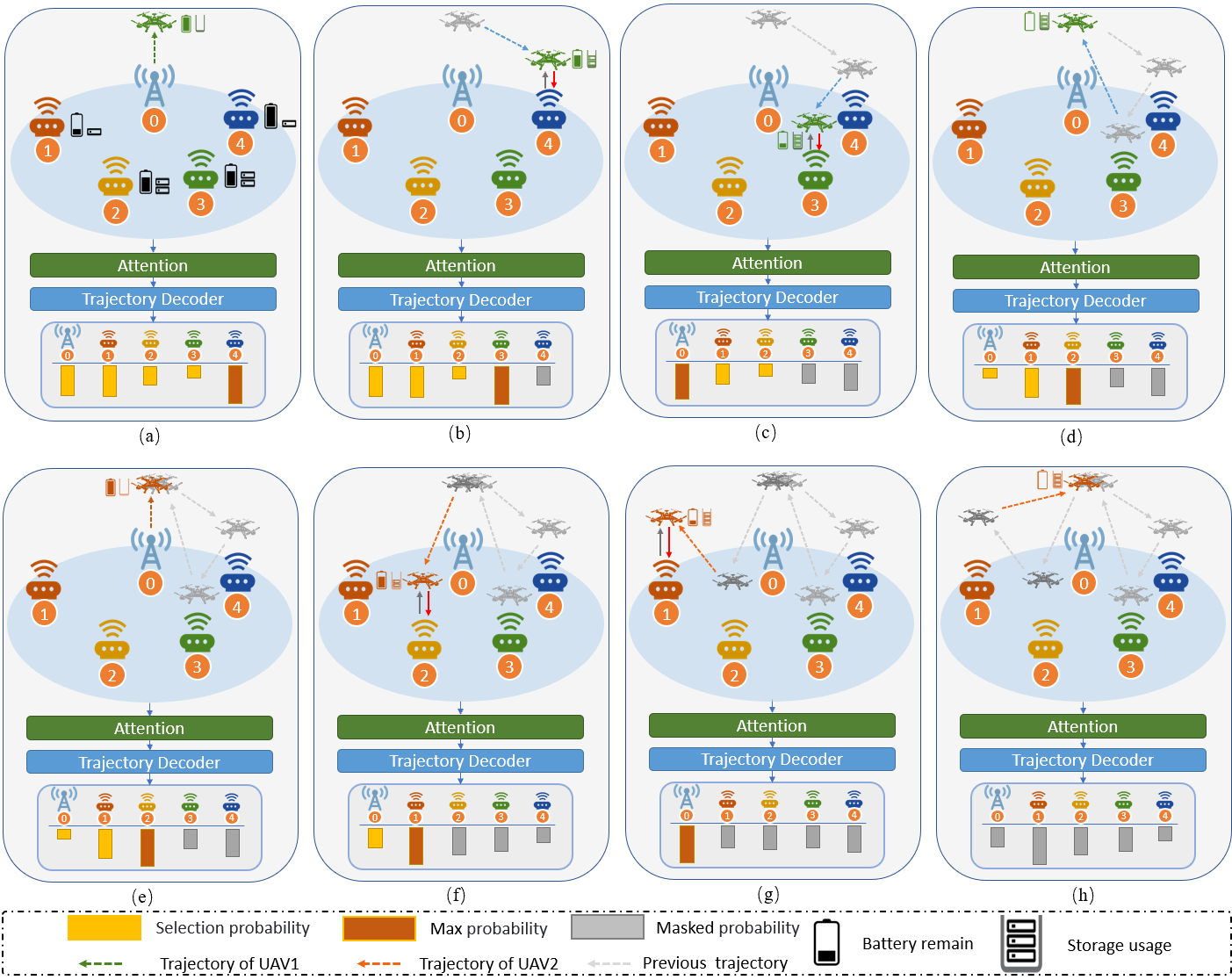}
\caption{The trajectories generated by the trajectory decoder. }
\label{fig:decoder}
\end{figure*}

\subsubsection{Graph State Definition}
We develop the graph state for the trajectory decoder that considers the constraints of UAV battery and storage capacity in every step of UAV trajectory planning. Therefore, the trajectory can be optimized under all constraints. 
In the $t$-th time step, the graph state is denoted as $\textbf{h}_s(t)$:
 \begin{equation}
\label{eq:scenario_state}
\textbf{h}_s(t)=[\textbf{h}_{sa}, \textbf{h}^L_{\pi_{t-1}}, C_{t}, E_{t}]
\end{equation}
where $\textbf{h}_{sa}$ is the graph feature obtained from the graph encoder, $\textbf{h}^L_{\pi_{t-1}}$ is the self-attention feature of the last selected IoTD in the trajectory of the current UAV, and $\pi_{t-1}$ is the index of the last selected IoTD. $C_{t}$ and $E_{t}$ represent the remaining data and battery capacity of the current UAV in the $t$-th time step, respectively.

\subsubsection{Trajectory Generation}
In each time step, we use the graph state $\textbf{h}_s(t)$ and the self-attention features of the $L$-th self attention layer to calculate the alignment vector as follows:
\begin{equation}\label{eq:a4}
	a_i\left(t\right)=\frac{\exp(\mathbf{h}_s\left(t\right)^T\mathbf{W}_{i}\mathbf{h}^L_{i})}{\sum_{j=1}^{N}{\exp({\mathbf{h}}_s(t)^T\mathbf{W}_{j}\mathbf{h}^L_{j})}}
\end{equation}
where $\mathbf{W}_{i}$ and $\mathbf{W}_{j}$ are learnable matrices.

Then, we can calculate the context vector as follows:
\begin{equation}\label{eq:a5}
	\mathbf{c}\left(t\right)=\sum_{i=1}^{N}{a_i\left(t\right)}\mathbf{h}^L_{i}
\end{equation}

Next, the probability distribution of the remaining IoTDs  $\textbf{P}(t)=\left\{p_i(t),j\in \mathcal{N} \right\}$ can be calculated as
\begin{align}
	\textbf{P}\left(t\right) &= \operatorname{softmax}\left(\mathbf{W}_{p}\tanh\left(\mathbf{W}_{c}\cdot\operatorname{Concat}\left(\mathbf{c}\left(t\right),\mathbf{h}_s\left(t\right)\right)\right)\right) \notag \\
	&\quad \cdot \textbf{M}
	\label{eq:a6}
\end{align}
where $\mathbf{W}_{p}$ and $\mathbf{W}_{c}$ are learnable matrices, $\textbf{M}$ is the mask matrix. $\textbf{M}(i)=1$ means UAV can collect data of the $i$-th IoTD, and $\textbf{M}(i)=0$ otherwise. $p_i(t)$ represents the probability that the current UAV selects the $i$-th IoTD as next IoTD to collect data in the $t$-th time step.

In each time step, the current UAV selects the IoTD with max probability in $\textbf{P}(t)$ as the next IoTD $\pi_t$ to collect data, so one has 
\begin{equation}
\label{eq:next}
\pi_t= \text{argmax} (\textbf{P}(t))
\end{equation}
where $\text{argmax}(\cdot)$ returns the \textcolor[rgb]{0,0,0}{index of the IoTD with the max probability. }

\subsubsection{Trajectory Segmentation}
\label{sec:split}
When all IoTDs have been selected, we can get the total trajectory $\boldsymbol{\pi}=[\pi_1, ..., \pi_T]$. $\boldsymbol{\pi}$ is the permutation of all IoTDs and the data center. $\pi_t \in \left\{ 1,...,N \right\}$ represents the IoTD index and $\pi_t=0$ represents the data center. For example in Fig. \ref{fig:decoder}, we can get the whole trajectory $\boldsymbol{\pi}=[0,4,3,0,2,1,0]$.
Then, the whole trajectory generated by trajectory decoder can be divided into several trajectories for all UAVs. For example, the trajectory $\boldsymbol{\pi}=[0,4,3,0,2,1,0]$ can be divided into $\boldsymbol{\pi}_{(1)}=[0,4,3,0]$ and $\boldsymbol{\pi}_{(2)}=[0,2,1,0]$. Therefore, the number of UAVs is set to $m=2$. Finally, all UAVs can fly and collect data in parallel according to the planned trajectories $\boldsymbol{\pi}_{(1)},...,\boldsymbol{\pi}_{(m)}$.


\subsection{Actor-Critic Trajectory Learning}
\label{sec:train}
The training method of the ATOM model is introduced in this section. 
In reinforcement learning, value-based methods, such as Q-learning and DQN, are inefficient for large-scale action space\cite{2019-np-alphago}. Traditional policy-based methods, such as vanilla policy gradient and Monte-Carlo policy gradient, are hard to converge \cite{jiang2021distributed}. 
Therefore, we utilize a novel TENMA method to train the ATOM model, in which an additional critic network is utilized to evaluate the ATOM model, and the real reward of the system is applied as the baseline to reduce the variance of the critic network, and enhance the stabilization and generalization of TENMA. The detailed procedure of the TENMA method can be described as follows:
\subsubsection{State, Action and Reward Definition} 
In the WPT-assisted IoT system, we define the state of the system as $\text{State}=\left\{ \textbf{h}_{sa}, \textbf{h}^L_{\pi_{t-1}}, C_{t}, E_{t} \right\}$. and then the action of UAVs is defined as the whole trajectory $\boldsymbol{\pi}$.

Moreover, we draw $K$ instances from the state space and use Monte Carlo simulation to produce feasible sequences with respect to the current policy of the ATOM model. The reward of the $k$-th instance is defined as
\begin{equation}
\label{eq:reward}
R_k=-\sum_{j=1}^m \sum_{t=1}^{s_j} ||r_j[\pi_{k,t+1}]-r_j[\pi_{k,t}]||_2
\end{equation}
\subsubsection{Actor Network Design}
The ATOM model with parameter $\theta_\pi$ is designed as the actor network, which defines a stochastic policy $P_{\theta_\pi} (\boldsymbol{\pi}_k \mid \textbf{s}_k)$  for selecting the trajectory $\boldsymbol{\pi}_k$ with the $k$-th instance as follows:
\begin{equation}
\label{eq:stochastic}
P_{\theta_\pi} (\boldsymbol{\pi}_k \mid \textbf{s}_k)=\prod_{j=1}^{m}\prod_{t=1}^{s_j} P_{\theta_\pi} (\pi_{k,t} \mid \textbf{s}_k, \boldsymbol{\pi}_{k,1:t-1})
\end{equation}
where $\textbf{s}_k$ is the state of the $k$-th instance. The loss function of the actor network is defined as
\begin{equation}	
	\label{eq:actorloss}
L\left(\theta_\pi\right)=\frac{1}{K}\sum_{k=1}^{K}{{-R}_k\log P_{\theta_\pi} (\boldsymbol{\pi}_k \mid \textbf{s}_k)}
\end{equation}

Finally, we optimize $L\left(\theta_\pi\right)$ using the following policy gradient with baseline\cite{1992reinforce}: 
\begin{equation}
	\label{eq:actorloss2}
d\theta_\pi\gets\frac{1}{K}\sum_{k=1}^{K}{\left(R_k-Q\left(\mathbf{s}_k,\boldsymbol{\pi}_k\middle|\theta_Q\right)\right)\nabla_{\theta_\pi}\log P_{\theta_\pi} (\boldsymbol{\pi}_k \mid \textbf{s}_k)}
\end{equation}
where $Q\left(\mathbf{s}_k, \boldsymbol{\pi}_k| \theta_Q\right)$ is the predicted reward approximation. 
\subsubsection{Critic Network Design}

We design a critic network with parameter $\theta_Q$ to predict the reward approximation for the $k$-th instance. The loss function of the critic network is defined as
\begin{equation}
	\label{eq:criticloss1}
L\left(\theta_Q\right)=\frac{1}{K} \sum_{k=1}^K\left(R_k-Q\left(\mathbf{s}_k, \boldsymbol{\pi}_k| \theta_Q\right)\right)^2
\end{equation}

The critic network is applied as a baseline to stabilize the learning process. Hence, we optimize $L\left(\theta_Q\right)$ using the following gradient estimator:
\begin{equation}
	\label{eq:criticloss2}
{d\theta}_Q\gets\frac{1}{K}\sum_{k=1}^{K}{\nabla_{\theta_Q}\left(R_k-Q(\mathbf{s}_k,\boldsymbol{\pi}_k|\theta_Q)\right)^2}
\end{equation}

The critic network is optimized in the direction of reducing the difference between the expected rewards and the true rewards during Monte Carlo rollouts. 

\subsection{Time Complexity Analysis of the ATOM model}

The time complexity of the ATOM model is calculated based on the graph encoder and the trajectory decoder. We analyze the time complexity of each part, then compute the overall time complexity of the ATOM model. According to the transformer model\cite{2017transformer}, the time complexity of graph encoder is $O(N^2 \times D  \times L)$, where $N$ is the current number of IoTDs, $D$ is the hidden dimension,  $L$ is the number of self-attention layers. The time complexity of the trajectory decoder is $O(N \times D^2)$. When the number of IoTDs $N$ is small, $D$ dominates the complexity of graph encoder and trajectory decoder. The bottleneck of the ATOM model thus lies in trajectory decoder. However, as the number of IoTDs grows larger, $N$ gradually dominates the complexity of these modules, in which case the graph encoder becomes the bottleneck of the ATOM model.

\section{Results and Discussion}
\label{sec:experiments}

\subsection{Parameter Settings}

\textcolor{black}{In the simulation, we consider an area of 1000 m $\times$ 1000 m, and the locations and sensory data of IoTDs are various in the scenario. We assume there are 500 IoTDs in the area. 
The data size $D_{i}$ for each IoTD is randomly selected in $[0.2,1.5]$ MB \cite{zhang2022joint}.  
Then, the detailed parameter settings of the AUTO framework are listed in Table \ref{tab:param} \cite{yu2021multi}.}
Moreover, we use PyTorch to implement the AUTO framework. All simulations are carried out in Python3.6 Environment running on Intel Xeon E5 CPU and NVIDIA Tesla T4 GPU with 32GB RAM.

\begin{table}[h]
	\caption{Parameters of the system model}
	\centering
	\setlength{\tabcolsep}{3mm}
	\renewcommand\arraystretch{1.25}
	\begin{tabular}{|c|c|}\hline
		Parameter&{Value}\\\hline
		
		Fly speed of UAV $v$ & 10 m/s\\
		Flight Height $H^F$ & 20 m\\
		Flight Power $P^F$&75 W\\
		Bandwidth $B$ & 2M Hz\\
		Transmitting power $P^{T}$ & 0.5 W\\
		Data collection power $P^{C}$ & 0.5 W\\
		Hover power $P^{H}$ & 50 W\\
		Noise power $\sigma^{2}$ & -110 dBm\\
		Number of Self-Attention layers $L$&3\\
		Number of heads $H$&8\\
		Hidden dimension of feed-forward sublayer&512\\
		\textcolor{black}{Learning rate decay}&0.98\\
		\textcolor{black}{Training epoch} &500\\
		\textcolor{black}{Experience pool capacity}&10000\\
		\textcolor{black}{Batch size}&100\\
	\textcolor{black}{Learning rate for the Critic network}&1e-3\\
	\textcolor{black}{Learning rate for the Actor network}&1e-5\\
		\hline
	\end{tabular}
	
	\label{tab:param}
\end{table}


\subsection{Comparison of Attention Mechanisms with Graph Pooling}
This experiment is presented to evaluate the performance of different attention mechanisms with various pooling operators.
The ATOM model is compared with two attention mechanisms: Luong attention (Luong-Attn)\cite{2015luong} and Area attention (Area-Attn)\cite{2019area}, and three pooling operators mean, sum and max are also considered.
 The minimum energy costs (Min.), average energy costs (Ave.), and standard deviation (Std.) are shown in Table \ref{tab:perform}. 

\begin{table}[]
	\caption{Comparison of different attention mechanisms with different pooling operators}
	\centering
	\setlength{\tabcolsep}{3.5mm}
	\renewcommand\arraystretch{1.25}
	\begin{tabular}{|l|l|l|l|l|}
		\hline
		Attention Type& Pool operator  & Min. & Ave. & Std. \\ \hline
		\multirow{3}{*}{Luong-Attn\cite{2015luong}} &  Mean&  289.57 & 293.25 & 3.58 \\ \cline{2-5} 
		& Sum & 291.15 & 295.58 & 4.32 \\ \cline{2-5} 
		& Max & 292.37 & 296.49 & 4.05 \\ \hline
		\multirow{3}{*}{Area-Attn\cite{2019area}} &  Mean& 262.26 & 267.28 &  4.77\\ \cline{2-5} 
		&  Sum& 263.74 & 268.69 & 4.65 \\ \cline{2-5} 
		&  Max& 263.35 & 268.27 & 4.61 \\ \hline
		\multirow{3}{*}{ATOM}
		& Mean & \textbf{262.75} & \textbf{264.32} & \textbf{1.39} \\ \cline{2-5} 
		&  Sum&262.72  & 265.48 & 2.54 \\ \cline{2-5} 
		&  Max& 263.02 &265.74  & 2.43 \\ \hline
	\end{tabular}
	\label{tab:perform}
\end{table}


\textcolor{black}{In Table \ref{tab:perform}, it is evident that the proposed ATOM model, which incorporates a multi-head self-attention layer with mean graph pooling in its graph encoder, consistently exhibits the lowest energy consumption across all scenarios. Several factors may contribute to this lower energy consumption: (1) Both luong-attn and area-atten are local attention mechanisms, whereas multi-head self-attention is a hybrid mechanism combining local and global attentions. Through multi-head self-attention, the graph encoder can capture local IoT characteristics of IoTDs at different scales as well as global characteristics of IoTDs, thereby achieving enhanced performance. (2) Graph embedding and pooling enhance attentional features from graph information. Particularly, mean pooling, compared to max pooling, preserves more information of IoTDs and is more sensitive to features of IoTDs than sum pooling, making it more suitable for our ATOM model.}

\subsection{Comparison of Trajectory Designers}
This experiment is used to compare the overall performance of the proposed AUTO framework with different trajectory designers in the WPT-assisted IoT system, and two different evaluations are presented. 

Firstly, we compare the energy cost and computing time of the ATOM model with 
Pointer Network (PN)\cite{bello2016neural}, Graph Neural Network (GNN)\cite{2017-cvrp-gnn} and Graph Pointer Network (GPN)\cite{2019-gpn-rl} using different reinforcement trainers (e.g., Reinforce, AC, A2C and the proposed TENMA)\cite{li2022deep}. The average energy costs are shown in Table \ref{tab:method}. 


%

\begin{table}[h]
	\caption{Average energy costs of different trajectory designers with different reinforcement trainers}
	\centering
	\setlength{\tabcolsep}{2.5mm}
	\renewcommand\arraystretch{1.25}
	\begin{tabular}{|p{35pt}<{\centering}|p{35pt}<{\centering}|p{35pt}<{\centering}|p{35pt}<{\centering}|p{35pt}<{\centering}|}\hline
		Method&{Reinforce}&{AC}&{A2C}&{TENMA}\\\hline
		PN\cite{bello2016neural}&287.88&287.42&285.04&285.83\\
		GNN\cite{2017-cvrp-gnn}&293.46&292.73&290.31&290.72\\
		GPN\cite{2019-gpn-rl}&275.48&273.25&272.85&274.46\\
	    ATOM&266.51&265.17&264.84&\textbf{264.32}\\
		\hline
	\end{tabular}
	\label{tab:method}
\end{table}

We can see from Table \ref{tab:method} that the energy cost of the AUTO framework (ATOM+TENMA) is the lowest. 
The reason can be interpreted by the following reasons: (1) The ATOM model with multi-head attention and graph operator is a powerful combinatorial optimization solver with high generalization. (2) The TENMA in the training process achieves the highest reward by minimizing the variance of the reward. 

\begin{figure}[htbp]
	\centering
	\includegraphics[width=9cm]{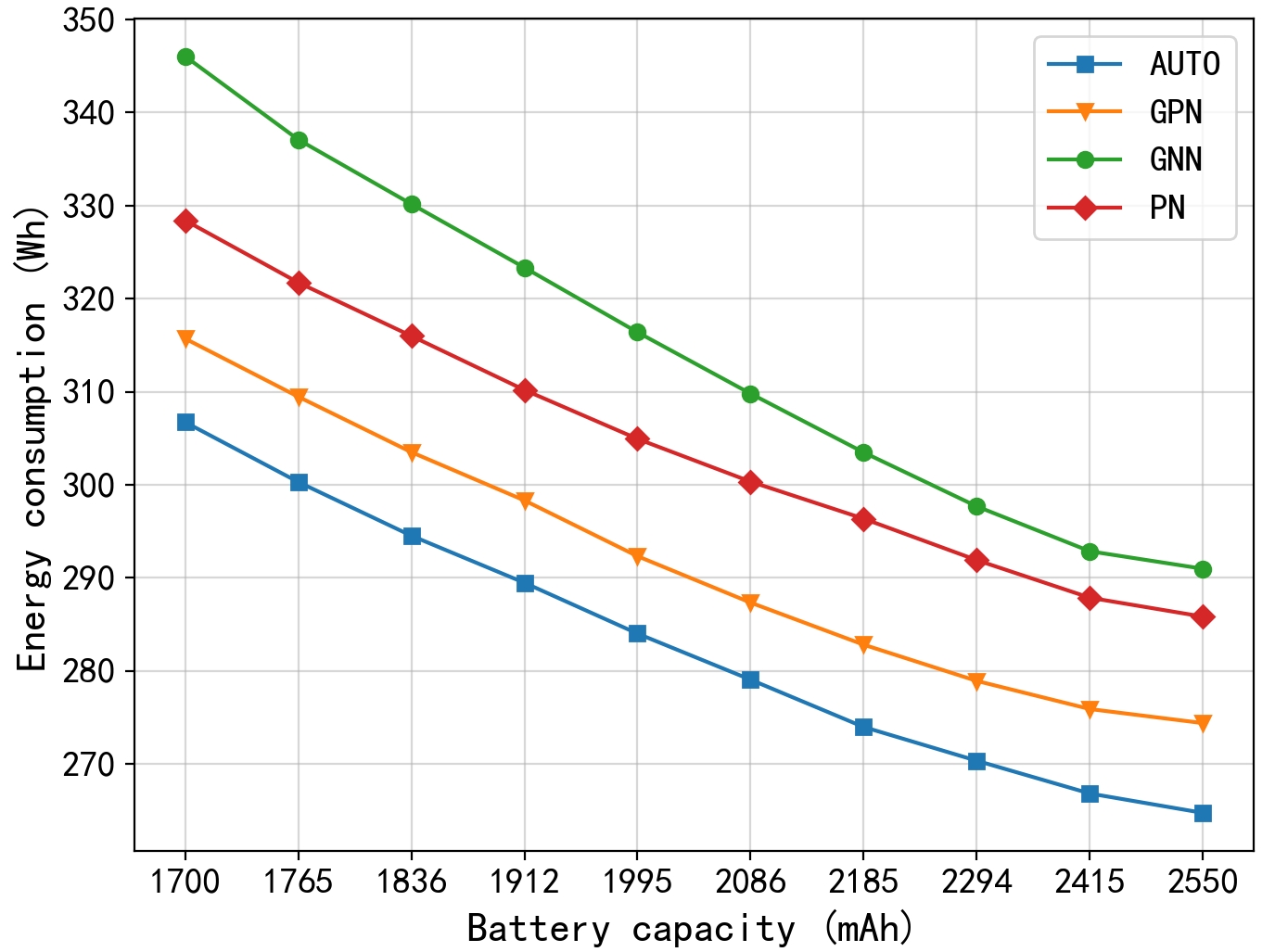}
	\caption{Energy cost with different battery capacities of UAVs.}
	\label{fig:change_energy}
\end{figure}

Secondly, to evaluate the generalization of the AUTO framework, we apply all neural network-based designers to scenarios with different UAV battery capacities. The range of UAV battery capacity is set from $1,700$ to $2,550$ mAh. The simulation results of the energy costs are shown in Fig. \ref{fig:change_energy}, in which we can see that GNN gets the highest energy cost, followed by PN, GPN, and AUTO. It is clear that the AUTO framework achieves the lowest energy cost among all neural network-based methods.

\subsection{Field Case Evaluation}

In this section, we design a field case to evaluate the performance of the AUTO framework in trajectory optimization. The simulation scenario is located at Yuelu Mountain in Hunan Province of China. We use $125$ IoTDs to monitor the air quality for three months. Each IoTD has an independent monitoring frequency, hence the amounts of sensory data collected in the IoTDs are different. The data center is located near the Changsha radio tower.

\begin{table}[h]
	\caption{DJI Matrice 100 specifications.}
	\centering
	\setlength{\tabcolsep}{6mm}
	\renewcommand\arraystretch{1.25}
	\begin{tabular}{|p{100pt}<{\centering}|p{70pt}<{\centering}|}\hline
		Parameter&{Value}\\\hline
		
		UAV type & DJI Matrice 100\\
		Hovering time & 40 min\\
		Max flight speed & 17 m/s\\
		Max power consumption & 350 W\\
		Battery capacity & 5700 mAh\\
		Max charging power & 180 W\\
		Flight control & Programmable\\
		Max data storage capability & 500 MB\\
		\hline
	\end{tabular}
	
	\label{tab:UAV}
\end{table}

\begin{figure}[htbp]
	\centering
	\includegraphics[width=8cm]{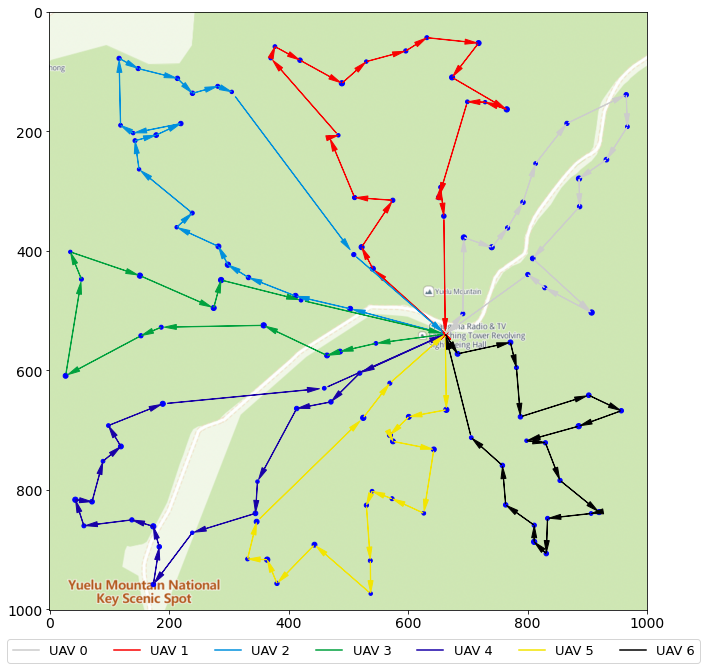}
	\caption{The generated trajectories of UAVs in the field case. }
	\label{fig:route}
\end{figure}

The programmable DJI Matrice 100 (M100) quad-copter drone with an open-source WPT hardware platform \cite{9184913} is achieved in the field case, whose specifications are summarized in Table \ref{tab:UAV}.
Fig. \ref{fig:route} shows the locations of $125$ IoTDs (blue dots) and data center (red cross). \textcolor{black}{The computation time of the trajectory planning is 1.04 s. The energy costs for different schemes are shown in Table \ref{tab:field}. Therefore, in this field case, the ATOM model can generate the best UAV trajectories, and all UAVs make maximum use of their batteries and data storage capacity.}
The trajectories of all UAVs are illustrated in Fig. \ref{fig:route}, 
in which the AUTO framework generates $7$ trajectories for UAVs. 

\begin{table}[h]
	\caption{The energy costs of different trajectory designers with different attention mechanisms}
	\centering
	\setlength{\tabcolsep}{2.5mm}
	\renewcommand\arraystretch{1.25}
	\begin{tabular}{|p{35pt}<{\centering}|p{55pt}<{\centering}|p{55pt}<{\centering}|p{40pt}<{\centering}|}\hline
		Method&{Luong-Atten}&{Area-Atten}&{MSA}\\\hline
		PN\cite{bello2016neural}&90.45&88.24&84.37\\
		GNN\cite{2017-cvrp-gnn}&98.77&98.23&92.14\\
		GPN\cite{2019-gpn-rl}&86.43&82.67&77.43\\
		ATOM&79.48&74.65&\textbf{72.73}\\
		\hline
	\end{tabular}
	\label{tab:field}
\end{table}

\section{Conclusions}
\label{sec:conclusion}

In this paper, we propose the AUTO framework to optimize the energy costs of UAVs in the WPT-assisted IoT system. In the AUTO framework, we first design the ATOM model based on graph transformer to generate the UAV trajectories with the lowest energy cost. In the ATOM model, the self-attention mechanism is utilized in graph encoder to extract the self-attention features of IoTDs. Then, the trajectory decoder generates the trajectories that can meet the resource constraints of UAVs. Next, we train the ATOM model by the TENMA method. 
Finally, the experiment results show that the performance of the AUTO framework surpasses other trajectory designers. 

\bibliographystyle{IEEEtran}

\bibliography{bare_jrnl}
\section*{Biographies}
\begin{IEEEbiography}[{\includegraphics[width=1in,height=1.25in,keepaspectratio]{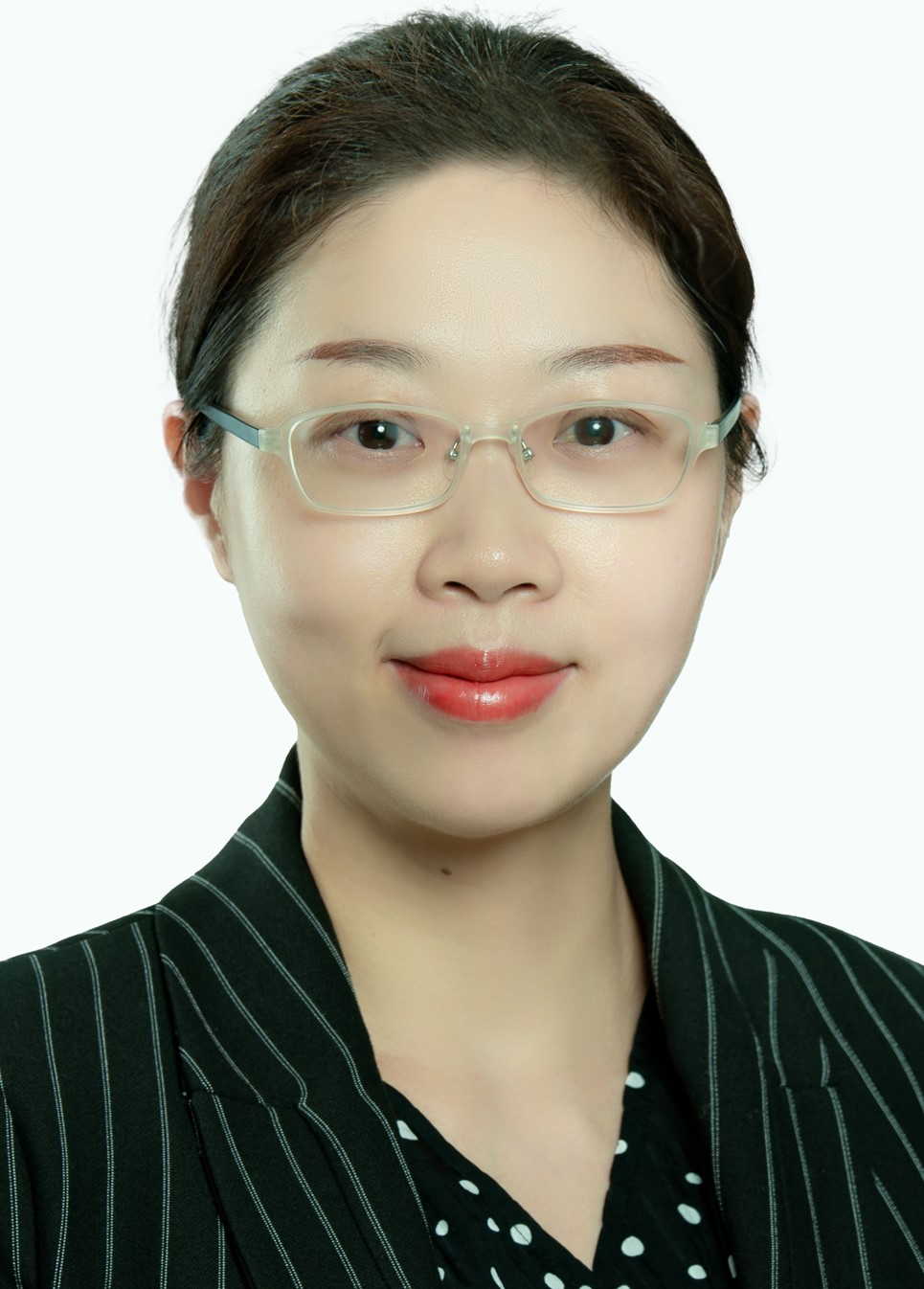}}]{Li Dong} received the B.S. and M.S. degrees in School of Physics and Electronics from Hunan Normal University, China, in 2004 and 2007, respectively. She received her Ph.D. degree in School of Geosciences and Info-physics from Central South University, China, in 2018. She is currently an associate professor at Hunan University of Technology and Business, China. Her research interests include Industrial Internet of Things, Machine Learning, and Mobile Edge Computing.
\end{IEEEbiography}
\vspace{-20 mm}
\begin{IEEEbiography}[{\includegraphics[width=1in,height=1.25in,keepaspectratio]{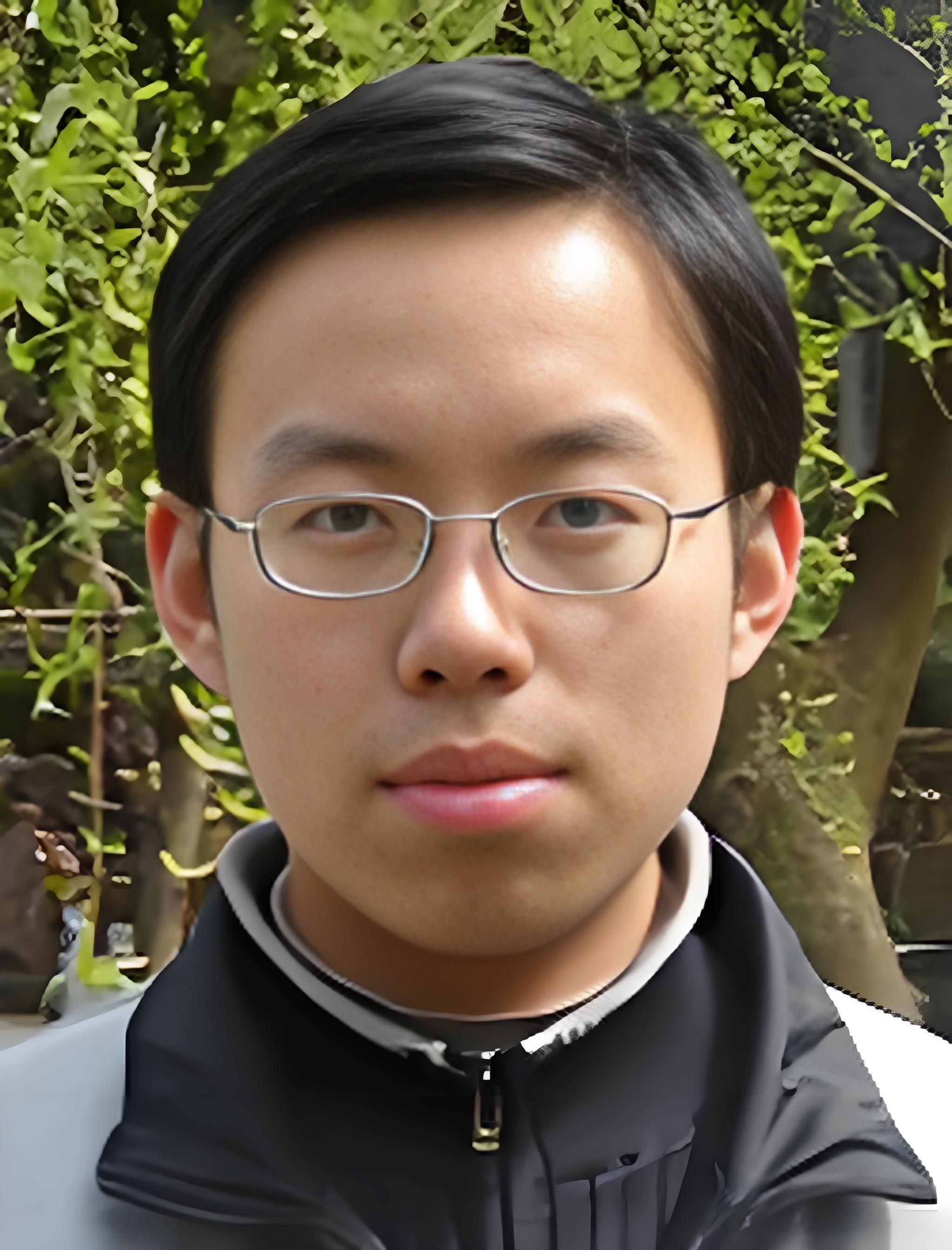}}]{Feibo Jiang} received his B.S. and M.S. degrees in School of Physics and Electronics from Hunan Normal University, China, in 2004 and 2007, respectively. He received his Ph.D. degree in School of Geosciences and Info-physics from Central South University, China, in 2014. He is currently an associate professor at the Hunan Provincial Key Laboratory of Intelligent Computing and Language Information Processing, Hunan Normal University, China. His research interests include  Industrial Internet of Things, Artificial Intelligence, and Mobile Edge Computing.
\end{IEEEbiography}
\vspace{-20 mm}
\begin{IEEEbiography}[{\includegraphics[width=1in,height=1.25in,keepaspectratio]{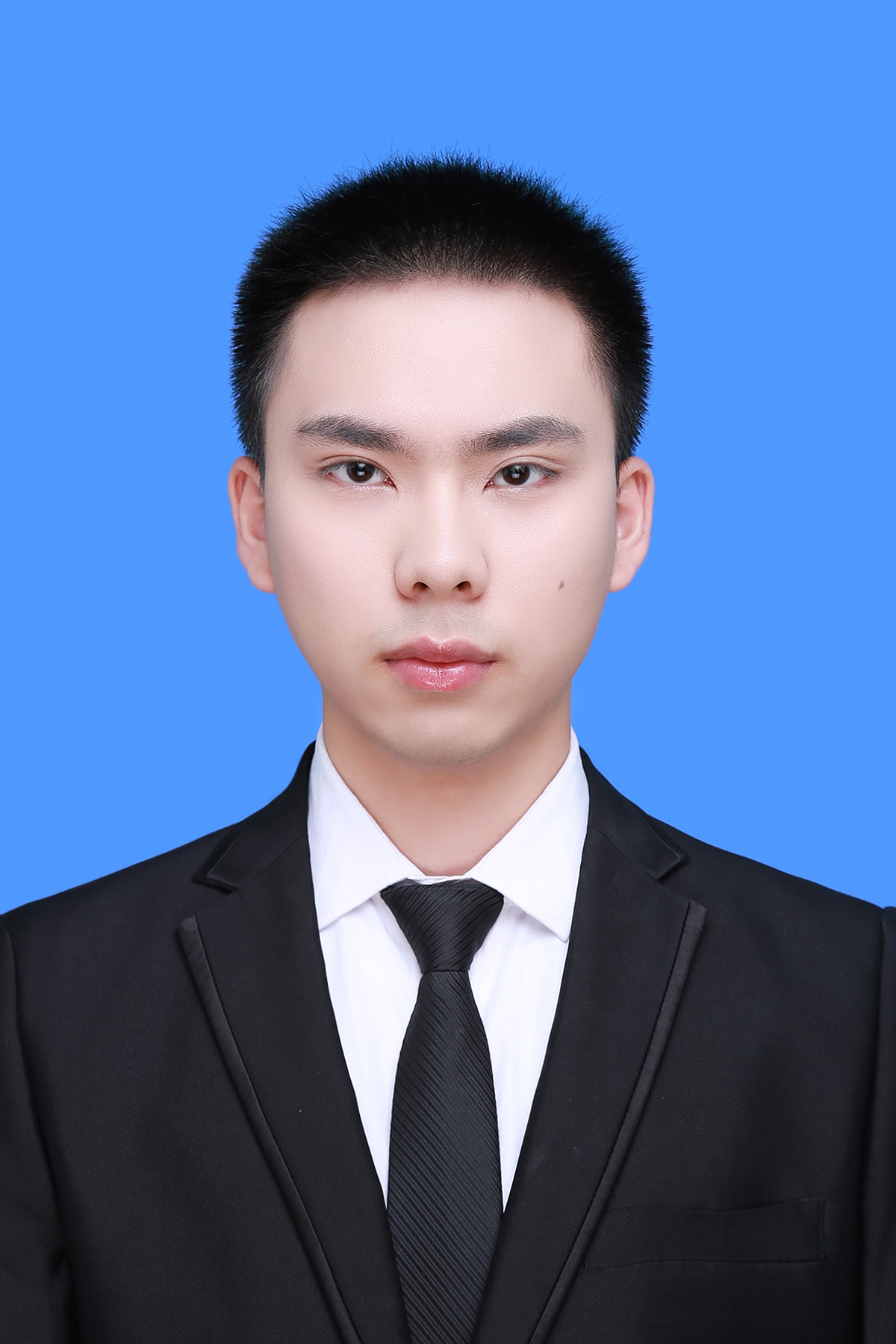}}]{Yubo Peng} received the B.S. and M.S. degrees in computer science and technology from Hunan Normal University, Changsha, China, in 2019 and 2024. He is currently working toward the doctor’s degree in software engineering from the School of Intelligent Software and Engineering, Nanjing University, Nanjing, China.
His main research interests include Federated Learning and Semantic Communication.
\end{IEEEbiography}
\vspace{-20 mm}
\newpage
\end{document}